\documentclass[aps,prl,reprint,twocolumn,floatfix,twoside,tbtags,showpacs]{revtex4-1}

\usepackage[utf8]{inputenc}
\usepackage{graphicx}
\usepackage{subfigure}
\usepackage{amsmath,amssymb}
\usepackage{bm}	
\usepackage{xspace} 
\usepackage[pdftex]{hyperref}

\newcommand{\abinitio}{\textit{ab initio}\@\xspace}
\newcommand{\Abinitio}{\textit{Ab initio}\@\xspace}

\begin{document}

\title{First-Principles Study of Secondary Slip in Zirconium}

\author{Nermine Chaari}
\affiliation{CEA, DEN, Service de Recherches de Métallurgie Physique, F-91191 Gif-sur-Yvette, France}

\author{Emmanuel Clouet}
\email{emmanuel.clouet@cea.fr}
\thanks{Corresponding author}
\affiliation{CEA, DEN, Service de Recherches de Métallurgie Physique, F-91191 Gif-sur-Yvette, France}

\author{David Rodney}
\affiliation{Institut Lumière Matière, Université Lyon 1, CNRS, UMR 5306, F-69622 Villeurbanne, France}

\date{17-10-2013}

\begin{abstract}
	Although the favored glide planes in hexagonal close-packed Zr are prismatic, screw dislocations can escape their habit plane to glide in either pyramidal or basal planes. Using \abinitio calculations within the nudged elastic band method, we show that, surprisingly, both events share the same thermally activated process with an unusual conservative motion of the prismatic stacking fault perpendicularly to itself. Halfway through the migration, the screw dislocation adopts a nonplanar metastable configuration with stacking faults in adjacent prismatic planes joined by a two-layer pyramidal twin. 
\end{abstract}

\maketitle

Plastic deformation in metals results mainly from the motion of line defects called dislocations. Many dislocation properties derive from the atomic-scale structure of their core, the region in the immediate vicinity of the dislocation line where crystallinity is disrupted. One such property is cross slip, i.e., the ability for screw dislocations to change glide plane, a stress-releaving process central to strain hardening and fatigue resistance \cite{Kubin2013, Friedel1964}.

Thus far, cross slip has mostly been studied in face-centered cubic (fcc) metals for the conventional planar dissociated $1/2 \langle 110 \rangle \{ 111 \}$ dislocations. Elasticity models \cite{Escaig1968, Friedel1964} confirmed by atomic-scale simulations \cite{Vegge2000, Rao2011} showed that the dominant cross-slip mechanism involves a local constriction of the dislocation in its initial glide plane followed by redissociation in the cross-slip plane (Friedel-Escaig mechanism). Another mechanism, which occurs under higher stresses met, for instance, in nanocrystalline plasticity \cite{Bitzek2008}, involves the successive change of glide plane of both partial dislocations \cite{Fleischer1959}. Mechanisms involving a metastable configuration of the screw dislocation spread over several planes are also possible, as found in iridium \cite{Cawkwell2005}. 

\begin{figure}[!tb]
\begin{center}
 \includegraphics[width=0.99 \linewidth]{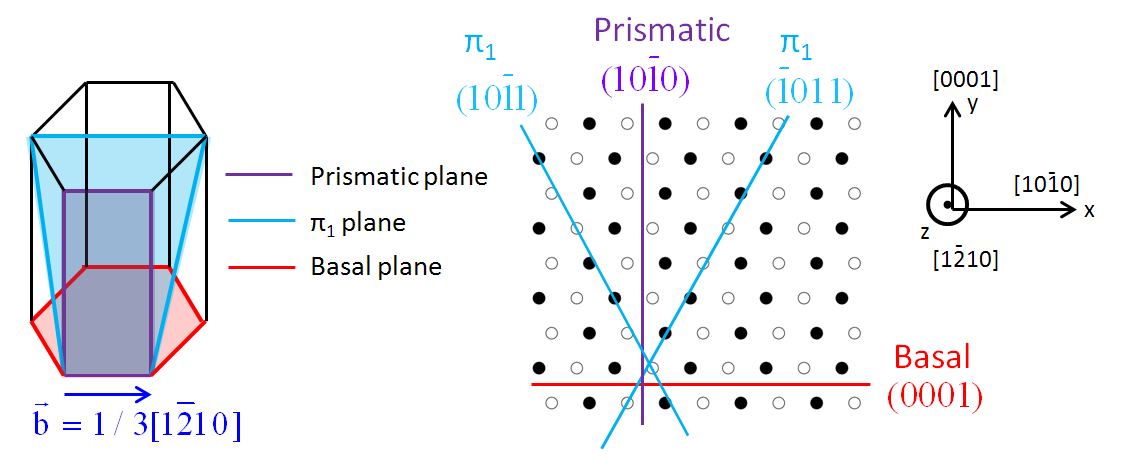} 
 \caption{Hexagonal close-packed structure showing the different potential glide planes for a screw dislocation of Burgers vector $\vec{b}=1/3[1\bar{2}10]$. A projection perpendicular to $\vec{b}$ is shown on the left, where atoms are sketched by circles with a color depending on their $(1\bar{2}10)$ plane.}
\label{figure1}
\end{center} 
\end{figure}

Cross slip is also observed in hexagonal close-packed (hcp) metals. In zirconium and titanium, $1/3\langle 1\bar{2}10 \rangle$ dislocations are dissociated and glissile in prismatic $\{10\bar{1}0\}$ planes \cite{Caillard2003}, as confirmed by first-principles calculations \cite{Ghazisaeidi2012, Clouet2012}. But screw dislocations have also been reported experimentally to glide at high temperatures in both first-order pyramidal $\pi_1$ $\{10\bar{1}1\}$ planes \cite{Shechtman1973, Naka1988, Rautenberg2012} and basal $\{0001\}$ planes \cite{Akhtar1973, Shechtman1973} (see Fig. \ref{figure1} for a graphical description of these planes). 
These secondary-slip processes do not correspond to cross slip as understood in fcc metals because the parent and the cross-slipped glide planes are not equivalent.
Rather, this secondary slip in hcp metals is related to the fundamental question: how can a dislocation dissociated in a plane glide in another plane?

In this Letter, we employ a combination of first-principles and empirical potential atomic-scale calculations to study secondary slip in hcp metals. We focus mainly on zirconium, although we checked that the present findings also apply to titanium. We show that unexpectedly, both basal and pyramidal slips are limited by the same thermally activated process, which involves an unusual motion of the prismatic stacking fault perpendicularly to itself, and results in a nonplanar metastable configuration composed of stacking faults in two successive prismatic planes joined by an elementary two-layer twin in the pyramidal plane.

We employed the \textsc{pwscf} code \cite{Giannozzi2009} as described in Ref. \cite{Clouet2012} to perform \abinitio calculations based on the density functional theory, an approach that has proved instrumental in unravelling dislocation core properties in various metals \cite{Ismail-Beigi2000, Woodward2002, Romaner2010, Pizzagalli2009a, Woodward2008, Clouet2012, Ghazisaeidi2012}. We also performed atomistic simulations with the embedded atom method (EAM) potential developed by Mendelev and Ackland \cite{Mendelev2007} to check the validity of our results in larger simulation cells. This potential is suitable to model screw dislocations, predicting, in particular, dissociation in a prismatic plane \cite{Khater2010}, in contrast with most other empirical potentials. The cells are fully periodic and contain a dislocation dipole with a periodic quadrupolar arrangement, described as an $S$ arrangement in a previous paper \cite{Clouet2012}. Since the two dislocations are equivalent, we consider below only one of them. The simulation cell has the following periodicity vectors: 
$\vec{a}_1 = n/2 \ a[10\bar{1}0]$, $\vec{a}_2 = m \ c[0001]$ and $\vec{a}_3 = \vec{b} = 1/3 \ a [1\bar{2}10]$, where $n$ and $m$ are two integers.

 \begin{figure}[!tbh]
\begin{center}
\includegraphics[width=0.99 \linewidth]{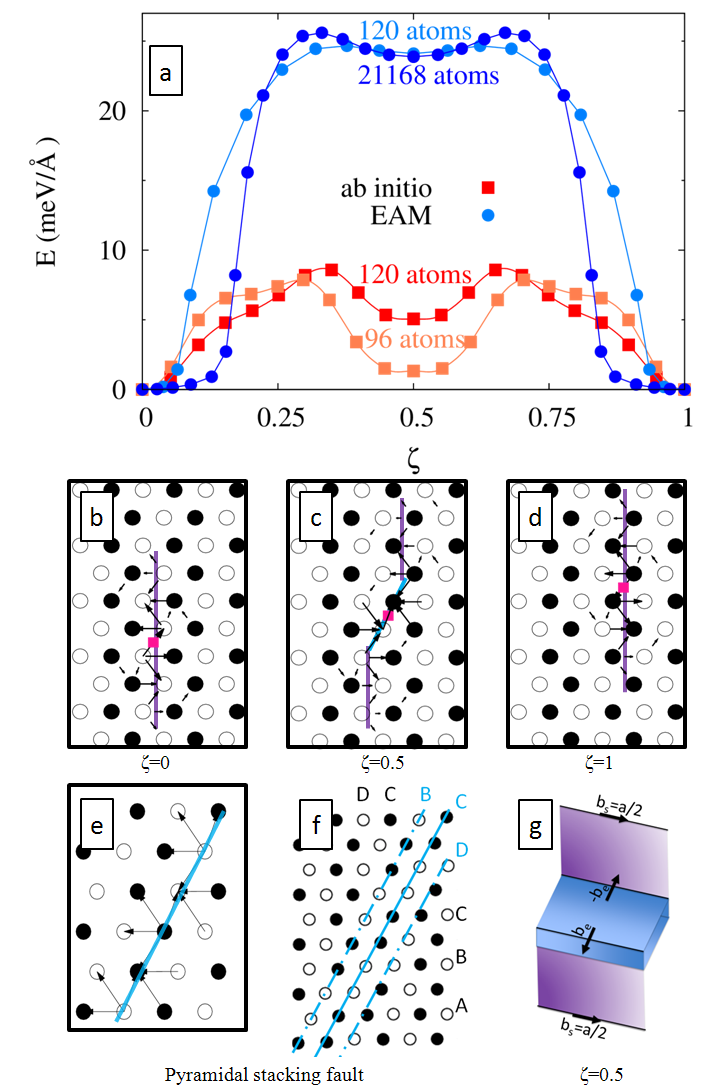} 
\caption{Secondary slip in the pyramidal $(\bar{1}011)$ plane.
(a) Energy barrier encountered by a screw dislocation dissociated in a prismatic plane when gliding in a pyramidal plane. Differential displacement maps show the dislocation  in its (b) initial, (c) intermediate, and (d) final states.
 The arrow between two atomic columns is proportional to the relative displacement of the columns in the $[1\bar{2}10]$ direction. 
 Displacements smaller than $0.08\,b$ are not shown.
 The pink squares indicate the dislocation center.
The pyramidal stacking fault is shown through its differential displacement map with respect to a perfect crystal in (e) and through its stacking sequence in (f), so as to display the two mirror planes ($B$ and $D$) associated with the twin structure. 
The intermediate dislocation configuration is sketched in (g).
 The prismatic and pyramidal faults are shown in blue and purple, respectively.}
 \label{figure2}
\end{center} 
\end{figure}

We first consider the glide in a pyramidal $\pi_1$ plane of a perfect screw dislocation dissociated in a prismatic plane [Fig. \ref{figure2}(b)]. To this end, we employ the nudged elastic band (NEB) method \cite{Henkelman2000} (with a tolerance on forces of 20\,meV/{\AA}) to compute the energy barrier against migration to the next equilibrium position along the $\pi_1$ plane, located at the intersection with the next prismatic plane [Fig. \ref{figure2}(d)]. The initial path was constructed by adding to the initial cut-plane of the dipole two new cut-planes that progressively shear inside a corrugated $\pi_1$ plane [plane in blue in Fig. \ref{figure2}(e)], thus resulting in the glide of the screw dislocations in these planes. We checked that other paths (for instance shearing in between two corrugated planes) lead to higher energy barriers. We also checked with the EAM potential that relaxed paths extending over multiple prismatic planes decompose into successions of elementary jumps following the mechanism described below. As shown in Fig. \ref{figure2}(a), both \abinitio and EAM potential calculations result in minimum energy paths with a local minimum at halfway, independently of the simulation cell size. 
\Abinitio calculations suffer from a size effect, but 
convergence is reached with the EAM potential for cells containing more than 2000 atoms. The difference between this local minimum and the energy of the initial configuration is estimated at about $\Delta E=3.2 \pm 1.6$\,meV/{\AA} with \abinitio calculations and $\Delta E = 24$\,meV/{\AA} with the EAM potential. 

The above intermediate state was found stable upon full relaxation after removal of the NEB constraint. The corresponding core structure is shown in  Fig. \ref{figure2}(c) by means of its differential displacement map \cite{Vitek1970}. The plastic strain of the dislocation spreads in the initial and final prismatic planes and in between, in a pyramidal $\pi_1$ plane. The dislocation center, shown as squares in Fig. \ref{figure2} and estimated from the symmetry of the differential displacements, lies in the $\pi_1$ plane, halfway between the initial and final prismatic planes. Ghazisaeidi and Trinkle \cite{Ghazisaeidi2012} obtained by \abinitio calculations a similar metastable configuration in titanium after relaxation of a dislocation initially introduced at the center position identified above, but the connection to nonprismatic glide was not made. On the other hand, the present NEB calculations show that this intermediate metastable core appears spontaneously in the course of pyramidal glide.

\begin{table}[!bth]
\begin{center}
\begin{tabular}{cccc}
\hline \hline
		&\multicolumn{2}{c}{Zr} & Ti    \\
		& \textsc{pwscf} &  EAM & \textsc{pwscf}     \\
\hline
Prismatic 	&  211		& 135	&  256      \\		
Pyramidal 	&  163  	& 243	&  227     \\		
\hline \hline
\end{tabular}
\caption{Energies in Zr and Ti of the stable stacking faults in the  $(10\bar{1}0)$ prismatic
and $(\bar{1}011)$ pyramidal  planes, 
corresponding, respectively, to the fault vectors $1/6 \ [1\bar{2}10]$
and $1/6 \ [1\bar{2}10] + b_e/\sqrt{3+4\gamma^2} \ [10\bar{1}2]$.
Energies are given in mJ\,m$^{-2}$.}
\label{tableau}
\end{center}
\end{table}

Spreading of the screw dislocation in the pyramidal $(\bar{1}011)$ plane results from the existence of a stable stacking fault in this plane. The fault appears when shearing the lattice in between two closely spaced atomic planes that form a corrugated pyramidal plane. Analyzing the atomic structure in the faulted region, we found that this fault corresponds to an elementary two-layer pyramidal twin, produced by the glide of a two-layer disconnection \cite{Serra1991}. The Burgers vector of the disconnection, which is also the fault vector, has a screw component equal to half a full screw dislocation $b_s=a/2$ and an edge component $b_e = a(4\gamma^2-9)/2\sqrt{3+4\gamma^2}$ ($\gamma$ is the $c/a$ ratio) \cite{Serra1991}. The structure of the elementary twin is displayed in Fig. \ref{figure2}(f), which is the same structure as in Fig. \ref{figure2}(e) but showing the atomic stacking rather than the differential displacements. In the pyramidal direction, the hcp structure corresponds to the stacking of corrugated planes {\dots}ABCDEFG\dots with no repeatable sequence. The glide of the two-layer disconnection creates locally a stacking {\dots}ABC$\mathit{\underline{D}C\underline{B}}$CDE{\dots}, with two mirror planes underlined in the previous sequence, $\mathit{\underline{D}}$ and $\mathit{\underline{B}}$.

\Abinitio calculations show that this pyramidal stacking fault has an energy lower than the prismatic stacking fault (see Table \ref{tableau}). The EAM potential leads to the reverse order, which is an artifact partially due to the low prismatic stacking fault predicted by this potential. Incidentally, this difference in fault energies also explains the higher energy barriers obtained in Fig. \ref{figure2} with the EAM potential. We also performed \abinitio calculations in titanium and found a similar stable pyramidal stacking fault with an energy still lower than the prismatic stacking fault (see Table \ref{tableau}).

A comparison of the pattern of differential displacements in the pyramidal stacking fault (\ref{figure2}e) and in the metastable configuration [Fig. \ref{figure2}(c)] shows that the fault corresponds to the section of the metastable core spread in the pyramidal plane. In the metastable core, the elementary twin is of finite length and is thus bordered by a dipole of disconnections. Since the disconnections have the same screw component ($b_s$) as the prismatic stacking fault, only the edge component ($b_e$) remains at the intersections. As illustrated in Fig. \ref{figure2}(g),
the metastable core may thus be described as two screw partial dislocations ($b_s$) in adjacent prismatic planes 
separated by two prismatic stacking faults and a two-layer pyramidal twin.
The prismatic faults and the twin are connected by stair rods forming a dipole of edge disconnections ($\pm b_e$).
The corresponding decomposition of the total Burgers vector is then
\begin{multline*}
	\frac{1}{3} \ a[1\bar{2}10] \to \frac{1}{6} \ a [1\bar{2}10]
		+ \frac{b_e}{\sqrt{3+4\gamma^2}}[10\bar{1}2] \\
		- \frac{b_e}{\sqrt{3+4\gamma^2}}[10\bar{1}2]
		+ \frac{1}{6} \ a [1\bar{2}10].
\end{multline*}

The dislocation glide mechanism in the $\pi_1$ plane may be described as a two-step process. The dislocation is initially dissociated into two partial dislocations of Burgers vector $b_s$ separated by a planar prismatic stacking fault [Fig. \ref{figure2}(b)]. In the first half of the minimum energy path, one partial dislocation and part of the stacking fault move perpendicularly to the habit plane and migrate to the next prismatic plane, creating the two-layer twin along a $\pi_1$ plane and resulting in the metastable dislocation configuration [Fig. \ref{figure2}(c)]. In the second half of the path, the remaining partial dislocation repeats the same process, removing the twin and restoring a planar fault spread in the final prismatic plane [Fig. \ref{figure2}(d)].
We should note that the partials, being of screw character, do not need to decompose and can glide out of the initial prismatic plane taking with them part of the stacking fault. We believe our calculations show the first instance of a conservative motion of a stacking fault perpendicularly to itself. 
The mechanism involves a collective motion of the atoms near the fault \footnote{See Supplemental Material at http://link.aps.org/supplemental/10.1103/PhysRevLett.112.075504 for movies showing the whole process for both pyramidal and basal slips.}. 
  
\begin{figure}[!tbh]
\begin{center}
\includegraphics[width=0.99 \linewidth]{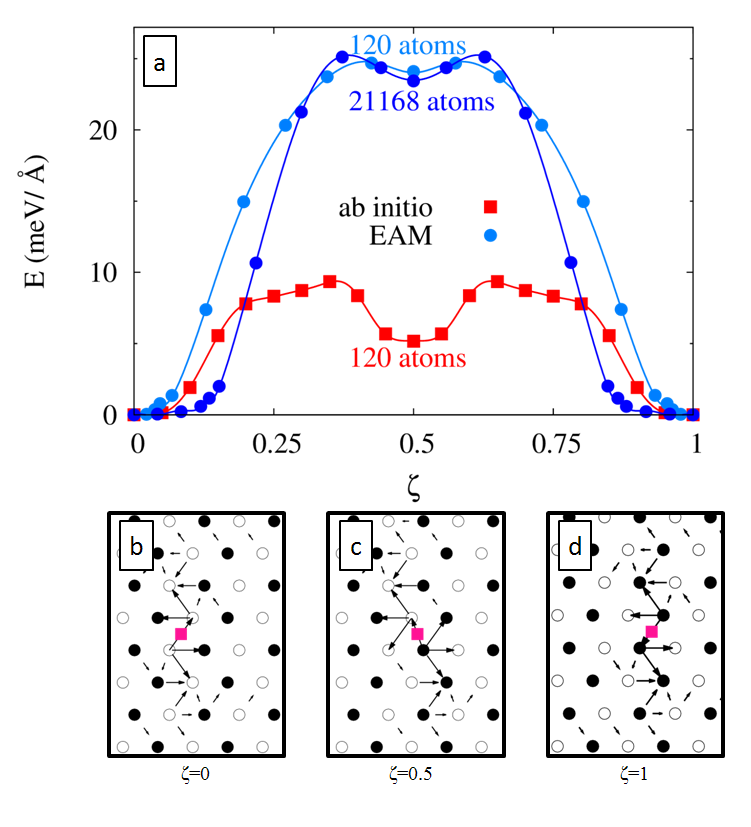} 
 \caption{Secondary slip in the basal $(0001)$ plane.
 (a) Energy barrier encountered by a screw dislocation when gliding in a basal plane. 
 The differential displacement maps show the dislocation (b) initial, (c) intermediate, and (d) final configurations.}
 \label{figure3}
\end{center} 
\end{figure}

Following the same approach, we investigated basal slip. We computed the energy barrier required for a screw dislocation spread in a prismatic plane to glide in a basal plane [Fig. \ref{figure3}(a)]. Surprisingly, we obtained the same energy barrier as for pyramidal slip and the local minimum at halfway ($\zeta=0.5$) corresponds to the same metastable configuration as obtained for pyramidal slip. The configurations in Figs. \ref{figure3}(c) and \ref{figure2}(c) are two variants of the same configuration, with the pyramidal fault lying in the $(10\bar{1}1)$ plane in Figs. \ref{figure3}(c) and in the $(\bar{1}011)$ plane in Fig. \ref{figure2}(c).

To understand the correspondence between pyramidal and basal slips, we determined the position of the dislocation in the $(1\bar{2}10)$ plane along both minimum energy paths. The position was deduced from the stress variation along the paths. Within elasticity theory \cite{Clouet2009b}, the total stress of a periodic unit cell containing a dislocation dipole of Burgers vector $\vec{b}$ is given by
\begin{equation}   
\sigma_{ij} =  C_{ijkl} \left( \varepsilon_{kl} - \dfrac{b_{k}A_{l} + b_{l}A_{k}}{2S} \right),  
\label{sigma}
\end{equation}
where $S$ is the area of the simulation cell perpendicular to the dislocation lines, $C_{ijkl}$ are the elastic constants of the perfect crystal, and $\varepsilon_{kl}$ is the homogeneous applied strain. The cut vector $\vec{A}$ defines the dislocation dipole and is directly related to the relative positions of the dislocations. For dislocations with a line direction along the $z$ axis, we have $A_x = y_1 - y_2$ and $A_y = x_2 - x_1$, where $(x_1,y_1)$ and $(x_2, y_2)$ are the coordinates of the dislocations with Burgers vector $\vec{b}$ and $-\vec{b}$, respectively. Hence, any variation of the dislocation positions results in a variation of the $\vec{A}$ vector, and in turn, a variation of the stress. Since the strain $\varepsilon_{kl}$ is kept constant along the paths, we can invert Eq. (\ref{sigma}) to determine the dislocation position from the stress variation along the minimum energy paths.
 
\begin{figure}[!tbh]
\begin{center}
\includegraphics[width=0.7 \linewidth]{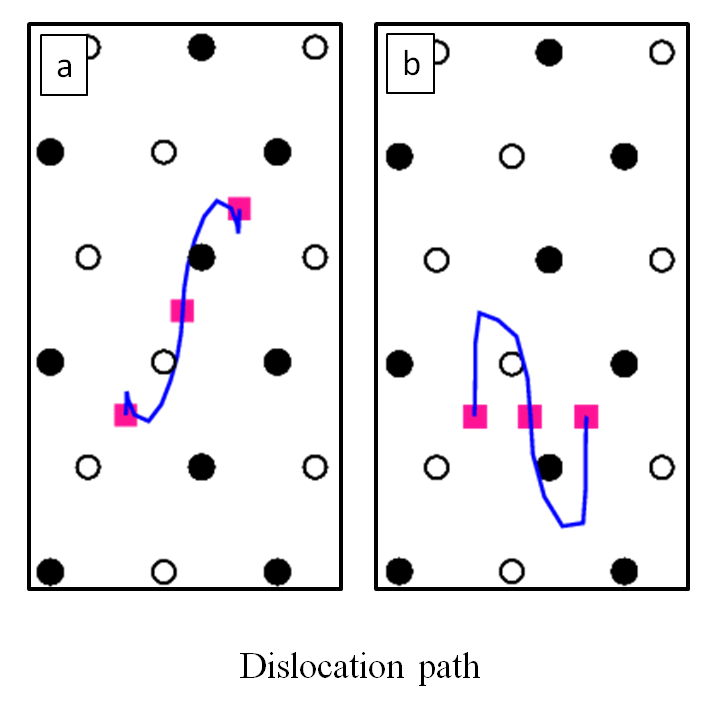} 
 \caption{Dislocation position in the $(1\bar{2}10)$ plane when slipping (a) in a pyramidal $\pi_1$ plane and (b) in a basal plane.
 The pink squares indicate the initial, intermediate and final positions.}
 \label{figure4}
\end{center} 
\end{figure} 	

We computed with the NEB method the same barriers as described previously but with the two dipole dislocations now moving in opposite directions to induce stress variations. The dislocation paths obtained with the EAM potential are shown in Fig. \ref{figure4}. For the initial ($\zeta=0$), intermediate ($\zeta=0.5$), and final states ($\zeta=1$), we found the same positions as previously determined from the differential displacement maps. For pyramidal slip [Fig. \ref{figure4}(a)], the dislocation takes the shortest path, which corresponds to a progressive shear of a $\pi_1$ corrugated plane. On the other hand, for basal slip [Fig. \ref{figure4}(b)], the path is not straight but can be decomposed in three steps : the dislocation glides first in a prismatic plane, then in a pyramidal plane, and finally again in a prismatic plane. Comparing Figs. \ref{figure4}(a) and \ref{figure4}(b), we see that the sections along the pyramidal planes are identical, apart from a symmetry operation between the two equivalent $(10\bar{1}1)$ and $(\bar{1}011)$ pyramidal planes. Since the energy barrier associated with prismatic glide is very low (less than 0.4\,meV\,\AA$^{-1}$ \cite{Clouet2012}), we now understand that pyramidal and basal slips are both limited by the same process, the elementary jump along the pyramidal plane, explaining why the same energy barrier is obtained for both pyramidal and basal slips.

Finally, to confirm the analogy between pyramidal and basal slips, we checked that they both have the same Peierls stress, defined as the critical shear stress at which the energy barrier disappears. To do so, a strain, corresponding to the target stress, was applied to the simulation cell and the energy barrier was recalculated using the NEB method. The two dislocations of the dipole were moved again in opposite directions to ensure that their motion agreed with their respective Peach-Koehler force directions. Calculations were performed only with the EAM potential in order to use a simulation cell large enough to neglect the variation of the elastic energy along the path. The only nonzero component of the applied stress tensor was $\tau_{yz}$ (see axis definition in Fig. \ref{figure1}). No $\tau_{xz}$ component could be applied because the Peierls stress for prismatic glide is so low (22\,MPa \cite{Khater2010}) that any $\tau_{xz}$ stress will cause the dislocations to glide in their prismatic plane. As expected, the energy barrier for pyramidal and basal slip canceled at the same applied shear stress, $\tau_{yz}=1.79$\,GPa. Since the EAM potential overestimates the energy barrier, we expect a lower Peierls stress with \abinitio calculations but the value will probably remain too high to allow for pyramidal and basal slip without thermal activation, in agreement with experiments where cross slip is observed only at high temperatures \cite{Akhtar1973, Shechtman1973, Naka1988, Rautenberg2012}.
   
To summarize, we have shown that pyramidal and basal slips share the same dislocation mechanism, 
involving a displacement of the prismatic stacking fault perpendicular to its habit plane and resulting in a metastable configuration partially spread in a pyramidal $\pi_1$ plane. This configuration has been found in both zirconium and titanium and involves the formation of an elementary two-layer twin inside the dislocation core. No constriction of the dislocation is necessary, in analogy with the Fleischer cross-slip mechanism in fcc metals. We anticipate that this new slip mechanism, which allows a dislocation spread in a plane to glide in another plane, will be of importance not only in hcp transition metals, but also in other more complex materials such as complex alloys or minerals.

\begin{acknowledgments}
	The authors thank Alexandre Prieur for the \abinitio calculations in Ti.
	This work was performed using HPC resources from GENCI-CINES, GENCI-CCRT and GENCI-IDRIS
  (Grants No. 2013-096847).
  The authors also acknowledge PRACE for awarding access to the Marenostrum resources 
  based in the Barcelona Supercomputing Center (project DIMAIM).
\end{acknowledgments}

\bibliographystyle{apsrev4-1}
\bibliography{article}

\begin{thebibliography}{28}%
\makeatletter
\providecommand \@ifxundefined [1]{%
 \@ifx{#1\undefined}
}%
\providecommand \@ifnum [1]{%
 \ifnum #1\expandafter \@firstoftwo
 \else \expandafter \@secondoftwo
 \fi
}%
\providecommand \@ifx [1]{%
 \ifx #1\expandafter \@firstoftwo
 \else \expandafter \@secondoftwo
 \fi
}%
\providecommand \natexlab [1]{#1}%
\providecommand \enquote  [1]{``#1''}%
\providecommand \bibnamefont  [1]{#1}%
\providecommand \bibfnamefont [1]{#1}%
\providecommand \citenamefont [1]{#1}%
\providecommand \href@noop [0]{\@secondoftwo}%
\providecommand \href [0]{\begingroup \@sanitize@url \@href}%
\providecommand \@href[1]{\@@startlink{#1}\@@href}%
\providecommand \@@href[1]{\endgroup#1\@@endlink}%
\providecommand \@sanitize@url [0]{\catcode `\\12\catcode `\$12\catcode
  `\&12\catcode `\#12\catcode `\^12\catcode `\_12\catcode `\%12\relax}%
\providecommand \@@startlink[1]{}%
\providecommand \@@endlink[0]{}%
\providecommand \url  [0]{\begingroup\@sanitize@url \@url }%
\providecommand \@url [1]{\endgroup\@href {#1}{\urlprefix }}%
\providecommand \urlprefix  [0]{URL }%
\providecommand \Eprint [0]{\href }%
\providecommand \doibase [0]{http://dx.doi.org/}%
\providecommand \selectlanguage [0]{\@gobble}%
\providecommand \bibinfo  [0]{\@secondoftwo}%
\providecommand \bibfield  [0]{\@secondoftwo}%
\providecommand \translation [1]{[#1]}%
\providecommand \BibitemOpen [0]{}%
\providecommand \bibitemStop [0]{}%
\providecommand \bibitemNoStop [0]{.\EOS\space}%
\providecommand \EOS [0]{\spacefactor3000\relax}%
\providecommand \BibitemShut  [1]{\csname bibitem#1\endcsname}%
\let\auto@bib@innerbib\@empty
\bibitem [{\citenamefont {Kubin}(2013)}]{Kubin2013}%
  \BibitemOpen
  \bibfield  {author} {\bibinfo {author} {\bibfnamefont {L.~P.}\ \bibnamefont
  {Kubin}},\ }\href@noop {} {\emph {\bibinfo {title} {Dislocations, mesoscale
  simulations and plastic flow}}},\ \bibinfo {edition} {first edition}\ ed.\
  (\bibinfo  {publisher} {Oxford University Press},\ \bibinfo {address}
  {Oxford, UK},\ \bibinfo {year} {2013})\BibitemShut {NoStop}%
\bibitem [{\citenamefont {Friedel}(1964)}]{Friedel1964}%
  \BibitemOpen
  \bibfield  {author} {\bibinfo {author} {\bibfnamefont {J.}~\bibnamefont
  {Friedel}},\ }\href@noop {} {\emph {\bibinfo {title} {Dislocations}}}\
  (\bibinfo  {publisher} {Pergamon Press},\ \bibinfo {address} {Oxford, UK},\
  \bibinfo {year} {1964})\BibitemShut {NoStop}%
\bibitem [{\citenamefont {Escaig}(1968)}]{Escaig1968}%
  \BibitemOpen
  \bibfield  {author} {\bibinfo {author} {\bibfnamefont {B.}~\bibnamefont
  {Escaig}},\ }\href {\doibase 10.1051/jphys:01968002902-3022500} {\bibfield
  {journal} {\bibinfo  {journal} {J. Phys. Paris}\ }\textbf {\bibinfo {volume}
  {29}},\ \bibinfo {pages} {225} (\bibinfo {year} {1968})}\BibitemShut
  {NoStop}%
\bibitem [{\citenamefont {Vegge}\ \emph {et~al.}(2000)\citenamefont {Vegge},
  \citenamefont {Rasmussen}, \citenamefont {Leffers}, \citenamefont
  {Pedersen},\ and\ \citenamefont {Jacobsen}}]{Vegge2000}%
  \BibitemOpen
  \bibfield  {author} {\bibinfo {author} {\bibfnamefont {T.}~\bibnamefont
  {Vegge}}, \bibinfo {author} {\bibfnamefont {T.}~\bibnamefont {Rasmussen}},
  \bibinfo {author} {\bibfnamefont {T.}~\bibnamefont {Leffers}}, \bibinfo
  {author} {\bibfnamefont {O.~B.}\ \bibnamefont {Pedersen}}, \ and\ \bibinfo
  {author} {\bibfnamefont {K.~W.}\ \bibnamefont {Jacobsen}},\ }\href {\doibase
  10.1103/PhysRevLett.85.3866} {\bibfield  {journal} {\bibinfo  {journal}
  {Phys. Rev. Lett.}\ }\textbf {\bibinfo {volume} {85}},\ \bibinfo {pages}
  {3866} (\bibinfo {year} {2000})}\BibitemShut {NoStop}%
\bibitem [{\citenamefont {Rao}\ \emph {et~al.}(2011)\citenamefont {Rao},
  \citenamefont {Dimiduk}, \citenamefont {Parthasarathy}, \citenamefont
  {El-Awady}, \citenamefont {Woodward},\ and\ \citenamefont {Uchic}}]{Rao2011}%
  \BibitemOpen
  \bibfield  {author} {\bibinfo {author} {\bibfnamefont {S.}~\bibnamefont
  {Rao}}, \bibinfo {author} {\bibfnamefont {D.}~\bibnamefont {Dimiduk}},
  \bibinfo {author} {\bibfnamefont {T.}~\bibnamefont {Parthasarathy}}, \bibinfo
  {author} {\bibfnamefont {J.}~\bibnamefont {El-Awady}}, \bibinfo {author}
  {\bibfnamefont {C.}~\bibnamefont {Woodward}}, \ and\ \bibinfo {author}
  {\bibfnamefont {M.}~\bibnamefont {Uchic}},\ }\href {\doibase
  10.1016/j.actamat.2011.08.029} {\bibfield  {journal} {\bibinfo  {journal}
  {Acta Mater.}\ }\textbf {\bibinfo {volume} {59}},\ \bibinfo {pages} {7135}
  (\bibinfo {year} {2011})}\BibitemShut {NoStop}%
\bibitem [{\citenamefont {Bitzek}\ \emph {et~al.}(2008)\citenamefont {Bitzek},
  \citenamefont {Brandl}, \citenamefont {Derlet},\ and\ \citenamefont
  {Van~Swygenhoven}}]{Bitzek2008}%
  \BibitemOpen
  \bibfield  {author} {\bibinfo {author} {\bibfnamefont {E.}~\bibnamefont
  {Bitzek}}, \bibinfo {author} {\bibfnamefont {C.}~\bibnamefont {Brandl}},
  \bibinfo {author} {\bibfnamefont {P.~M.}\ \bibnamefont {Derlet}}, \ and\
  \bibinfo {author} {\bibfnamefont {H.}~\bibnamefont {Van~Swygenhoven}},\
  }\href {\doibase 10.1103/PhysRevLett.100.235501} {\bibfield  {journal}
  {\bibinfo  {journal} {Phys. Rev. Lett.}\ }\textbf {\bibinfo {volume} {100}},\
  \bibinfo {pages} {235501} (\bibinfo {year} {2008})}\BibitemShut {NoStop}%
\bibitem [{\citenamefont {Fleischer}(1959)}]{Fleischer1959}%
  \BibitemOpen
  \bibfield  {author} {\bibinfo {author} {\bibfnamefont {R.}~\bibnamefont
  {Fleischer}},\ }\href {\doibase 10.1016/0001-6160(59)90122-1} {\bibfield
  {journal} {\bibinfo  {journal} {Acta Metall.}\ }\textbf {\bibinfo {volume}
  {7}},\ \bibinfo {pages} {134} (\bibinfo {year} {1959})}\BibitemShut {NoStop}%
\bibitem [{\citenamefont {Cawkwell}\ \emph {et~al.}(2005)\citenamefont
  {Cawkwell}, \citenamefont {Nguyen-Manh}, \citenamefont {Woodward},
  \citenamefont {Pettifor},\ and\ \citenamefont {Vitek}}]{Cawkwell2005}%
  \BibitemOpen
  \bibfield  {author} {\bibinfo {author} {\bibfnamefont {M.~J.}\ \bibnamefont
  {Cawkwell}}, \bibinfo {author} {\bibfnamefont {D.}~\bibnamefont
  {Nguyen-Manh}}, \bibinfo {author} {\bibfnamefont {C.}~\bibnamefont
  {Woodward}}, \bibinfo {author} {\bibfnamefont {D.~G.}\ \bibnamefont
  {Pettifor}}, \ and\ \bibinfo {author} {\bibfnamefont {V.}~\bibnamefont
  {Vitek}},\ }\href {\doibase 10.1126/science.1114704} {\bibfield  {journal}
  {\bibinfo  {journal} {Science}\ }\textbf {\bibinfo {volume} {309}},\ \bibinfo
  {pages} {1059} (\bibinfo {year} {2005})}\BibitemShut {NoStop}%
\bibitem [{\citenamefont {Caillard}\ and\ \citenamefont
  {Martin}(2003)}]{Caillard2003}%
  \BibitemOpen
  \bibfield  {author} {\bibinfo {author} {\bibfnamefont {D.}~\bibnamefont
  {Caillard}}\ and\ \bibinfo {author} {\bibfnamefont {J.~L.}\ \bibnamefont
  {Martin}},\ }\href {http://www.sciencedirect.com/science/bookseries/14701804}
  {\emph {\bibinfo {title} {Thermally activated mechanisms in crystal
  plasticity}}},\ edited by\ \bibinfo {editor} {\bibfnamefont {R.~W.}\
  \bibnamefont {Cahn}}\ (\bibinfo  {publisher} {Pergamon},\ \bibinfo {address}
  {Amsterdam},\ \bibinfo {year} {2003})\BibitemShut {NoStop}%
\bibitem [{\citenamefont {Ghazisaeidi}\ and\ \citenamefont
  {Trinkle}(2012)}]{Ghazisaeidi2012}%
  \BibitemOpen
  \bibfield  {author} {\bibinfo {author} {\bibfnamefont {M.}~\bibnamefont
  {Ghazisaeidi}}\ and\ \bibinfo {author} {\bibfnamefont {D.}~\bibnamefont
  {Trinkle}},\ }\href {\doibase 10.1016/j.actamat.2011.11.024} {\bibfield
  {journal} {\bibinfo  {journal} {Acta Mater.}\ }\textbf {\bibinfo {volume}
  {60}},\ \bibinfo {pages} {1287} (\bibinfo {year} {2012})}\BibitemShut
  {NoStop}%
\bibitem [{\citenamefont {Clouet}(2012)}]{Clouet2012}%
  \BibitemOpen
  \bibfield  {author} {\bibinfo {author} {\bibfnamefont {E.}~\bibnamefont
  {Clouet}},\ }\href {\doibase 10.1103/PhysRevB.86.144104} {\bibfield
  {journal} {\bibinfo  {journal} {Phys. Rev. B}\ }\textbf {\bibinfo {volume}
  {86}},\ \bibinfo {pages} {144104} (\bibinfo {year} {2012})}\BibitemShut
  {NoStop}%
\bibitem [{\citenamefont {Shechtman}\ and\ \citenamefont
  {Brandon}(1973)}]{Shechtman1973}%
  \BibitemOpen
  \bibfield  {author} {\bibinfo {author} {\bibfnamefont {D.}~\bibnamefont
  {Shechtman}}\ and\ \bibinfo {author} {\bibfnamefont {D.~G.}\ \bibnamefont
  {Brandon}},\ }\href {\doibase 10.1007/BF00549337} {\bibfield  {journal}
  {\bibinfo  {journal} {J. Mater. Sci.}\ }\textbf {\bibinfo {volume} {8}},\
  \bibinfo {pages} {1233} (\bibinfo {year} {1973})}\BibitemShut {NoStop}%
\bibitem [{\citenamefont {Naka}\ \emph {et~al.}(1988)\citenamefont {Naka},
  \citenamefont {Lasalmonie}, \citenamefont {Costa},\ and\ \citenamefont
  {Kubin}}]{Naka1988}%
  \BibitemOpen
  \bibfield  {author} {\bibinfo {author} {\bibfnamefont {S.}~\bibnamefont
  {Naka}}, \bibinfo {author} {\bibfnamefont {A.}~\bibnamefont {Lasalmonie}},
  \bibinfo {author} {\bibfnamefont {P.}~\bibnamefont {Costa}}, \ and\ \bibinfo
  {author} {\bibfnamefont {L.~P.}\ \bibnamefont {Kubin}},\ }\href {\doibase
  10.1080/01418618808209916} {\bibfield  {journal} {\bibinfo  {journal}
  {Philos. Mag. A}\ }\textbf {\bibinfo {volume} {57}},\ \bibinfo {pages} {717}
  (\bibinfo {year} {1988})}\BibitemShut {NoStop}%
\bibitem [{\citenamefont {Rautenberg}\ \emph {et~al.}(2012)\citenamefont
  {Rautenberg}, \citenamefont {Feaugas}, \citenamefont {Poquillon},\ and\
  \citenamefont {Cloué}}]{Rautenberg2012}%
  \BibitemOpen
  \bibfield  {author} {\bibinfo {author} {\bibfnamefont {M.}~\bibnamefont
  {Rautenberg}}, \bibinfo {author} {\bibfnamefont {X.}~\bibnamefont {Feaugas}},
  \bibinfo {author} {\bibfnamefont {D.}~\bibnamefont {Poquillon}}, \ and\
  \bibinfo {author} {\bibfnamefont {J.-M.}\ \bibnamefont {Cloué}},\ }\href
  {\doibase 10.1016/j.actamat.2012.04.001} {\bibfield  {journal} {\bibinfo
  {journal} {Acta Mater.}\ }\textbf {\bibinfo {volume} {60}},\ \bibinfo {pages}
  {4319} (\bibinfo {year} {2012})}\BibitemShut {NoStop}%
\bibitem [{\citenamefont {Akhtar}(1973)}]{Akhtar1973}%
  \BibitemOpen
  \bibfield  {author} {\bibinfo {author} {\bibfnamefont {A.}~\bibnamefont
  {Akhtar}},\ }\href {\doibase 10.1016/0001-6160(73)90213-7} {\bibfield
  {journal} {\bibinfo  {journal} {Acta Metall.}\ }\textbf {\bibinfo {volume}
  {21}},\ \bibinfo {pages} {1} (\bibinfo {year} {1973})}\BibitemShut {NoStop}%
\bibitem [{\citenamefont {Giannozzi}\ \emph {et~al.}(2009)\citenamefont
  {Giannozzi} \emph {et~al.}}]{Giannozzi2009}%
  \BibitemOpen
  \bibfield  {author} {\bibinfo {author} {\bibfnamefont {P.}~\bibnamefont
  {Giannozzi}} \emph {et~al.},\ }\href {\doibase
  10.1088/0953-8984/21/39/395502} {\bibfield  {journal} {\bibinfo  {journal}
  {J. Phys.: Condens. Matter}\ }\textbf {\bibinfo {volume} {21}},\ \bibinfo
  {pages} {395502} (\bibinfo {year} {2009})}\BibitemShut {NoStop}%
\bibitem [{\citenamefont {Ismail-Beigi}\ and\ \citenamefont
  {Arias}(2000)}]{Ismail-Beigi2000}%
  \BibitemOpen
  \bibfield  {author} {\bibinfo {author} {\bibfnamefont {S.}~\bibnamefont
  {Ismail-Beigi}}\ and\ \bibinfo {author} {\bibfnamefont {T.~A.}\ \bibnamefont
  {Arias}},\ }\href {\doibase 10.1103/PhysRevLett.84.1499} {\bibfield
  {journal} {\bibinfo  {journal} {Phys. Rev. Lett.}\ }\textbf {\bibinfo
  {volume} {84}},\ \bibinfo {pages} {1499} (\bibinfo {year}
  {2000})}\BibitemShut {NoStop}%
\bibitem [{\citenamefont {Woodward}\ and\ \citenamefont
  {Rao}(2002)}]{Woodward2002}%
  \BibitemOpen
  \bibfield  {author} {\bibinfo {author} {\bibfnamefont {C.}~\bibnamefont
  {Woodward}}\ and\ \bibinfo {author} {\bibfnamefont {S.~I.}\ \bibnamefont
  {Rao}},\ }\href {\doibase 10.1103/PhysRevLett.88.216402} {\bibfield
  {journal} {\bibinfo  {journal} {Phys. Rev. Lett.}\ }\textbf {\bibinfo
  {volume} {88}},\ \bibinfo {pages} {216402} (\bibinfo {year}
  {2002})}\BibitemShut {NoStop}%
\bibitem [{\citenamefont {Romaner}\ \emph {et~al.}(2010)\citenamefont
  {Romaner}, \citenamefont {Ambrosch-Draxl},\ and\ \citenamefont
  {Pippan}}]{Romaner2010}%
  \BibitemOpen
  \bibfield  {author} {\bibinfo {author} {\bibfnamefont {L.}~\bibnamefont
  {Romaner}}, \bibinfo {author} {\bibfnamefont {C.}~\bibnamefont
  {Ambrosch-Draxl}}, \ and\ \bibinfo {author} {\bibfnamefont {R.}~\bibnamefont
  {Pippan}},\ }\href {\doibase 10.1103/PhysRevLett.104.195503} {\bibfield
  {journal} {\bibinfo  {journal} {Phys. Rev. Lett.}\ }\textbf {\bibinfo
  {volume} {104}},\ \bibinfo {pages} {195503} (\bibinfo {year}
  {2010})}\BibitemShut {NoStop}%
\bibitem [{\citenamefont {Pizzagalli}\ \emph {et~al.}(2009)\citenamefont
  {Pizzagalli}, \citenamefont {Godet},\ and\ \citenamefont
  {Brochard}}]{Pizzagalli2009a}%
  \BibitemOpen
  \bibfield  {author} {\bibinfo {author} {\bibfnamefont {L.}~\bibnamefont
  {Pizzagalli}}, \bibinfo {author} {\bibfnamefont {J.}~\bibnamefont {Godet}}, \
  and\ \bibinfo {author} {\bibfnamefont {S.}~\bibnamefont {Brochard}},\ }\href
  {\doibase 10.1103/PhysRevLett.103.065505} {\bibfield  {journal} {\bibinfo
  {journal} {Phys. Rev. Lett.}\ }\textbf {\bibinfo {volume} {103}},\ \bibinfo
  {eid} {065505} (\bibinfo {year} {2009})}\BibitemShut {NoStop}%
\bibitem [{\citenamefont {Woodward}\ \emph {et~al.}(2008)\citenamefont
  {Woodward}, \citenamefont {Trinkle}, \citenamefont {{Hector, Jr.}},\ and\
  \citenamefont {Olmsted}}]{Woodward2008}%
  \BibitemOpen
  \bibfield  {author} {\bibinfo {author} {\bibfnamefont {C.}~\bibnamefont
  {Woodward}}, \bibinfo {author} {\bibfnamefont {D.~R.}\ \bibnamefont
  {Trinkle}}, \bibinfo {author} {\bibfnamefont {L.~G.}\ \bibnamefont {{Hector,
  Jr.}}}, \ and\ \bibinfo {author} {\bibfnamefont {D.~L.}\ \bibnamefont
  {Olmsted}},\ }\href {\doibase 10.1103/PhysRevLett.100.045507} {\bibfield
  {journal} {\bibinfo  {journal} {Phys. Rev. Lett.}\ }\textbf {\bibinfo
  {volume} {100}},\ \bibinfo {pages} {045507} (\bibinfo {year}
  {2008})}\BibitemShut {NoStop}%
\bibitem [{\citenamefont {Mendelev}\ and\ \citenamefont
  {Ackland}(2007)}]{Mendelev2007}%
  \BibitemOpen
  \bibfield  {author} {\bibinfo {author} {\bibfnamefont {M.~I.}\ \bibnamefont
  {Mendelev}}\ and\ \bibinfo {author} {\bibfnamefont {G.~J.}\ \bibnamefont
  {Ackland}},\ }\href {\doibase 10.1080/09500830701191393} {\bibfield
  {journal} {\bibinfo  {journal} {Philos. Mag. Lett.}\ }\textbf {\bibinfo
  {volume} {87}},\ \bibinfo {pages} {349} (\bibinfo {year} {2007})}\BibitemShut
  {NoStop}%
\bibitem [{\citenamefont {Khater}\ and\ \citenamefont
  {Bacon}(2010)}]{Khater2010}%
  \BibitemOpen
  \bibfield  {author} {\bibinfo {author} {\bibfnamefont {H.~A.}\ \bibnamefont
  {Khater}}\ and\ \bibinfo {author} {\bibfnamefont {D.~J.}\ \bibnamefont
  {Bacon}},\ }\href {\doibase 10.1016/j.actamat.2010.01.028} {\bibfield
  {journal} {\bibinfo  {journal} {Acta Mater.}\ }\textbf {\bibinfo {volume}
  {58}},\ \bibinfo {pages} {2978} (\bibinfo {year} {2010})}\BibitemShut
  {NoStop}%
\bibitem [{\citenamefont {Henkelman}\ and\ \citenamefont
  {Jónsson}(2000)}]{Henkelman2000}%
  \BibitemOpen
  \bibfield  {author} {\bibinfo {author} {\bibfnamefont {G.}~\bibnamefont
  {Henkelman}}\ and\ \bibinfo {author} {\bibfnamefont {H.}~\bibnamefont
  {Jónsson}},\ }\href {\doibase 10.1063/1.1323224} {\bibfield  {journal}
  {\bibinfo  {journal} {J. Chem. Phys.}\ }\textbf {\bibinfo {volume} {113}},\
  \bibinfo {pages} {9978} (\bibinfo {year} {2000})}\BibitemShut {NoStop}%
\bibitem [{\citenamefont {Vitek}\ \emph {et~al.}(1970)\citenamefont {Vitek},
  \citenamefont {Perrin},\ and\ \citenamefont {Bowen}}]{Vitek1970}%
  \BibitemOpen
  \bibfield  {author} {\bibinfo {author} {\bibfnamefont {V.}~\bibnamefont
  {Vitek}}, \bibinfo {author} {\bibfnamefont {R.~C.}\ \bibnamefont {Perrin}}, \
  and\ \bibinfo {author} {\bibfnamefont {D.~K.}\ \bibnamefont {Bowen}},\ }\href
  {\doibase 10.1080/14786437008238490} {\bibfield  {journal} {\bibinfo
  {journal} {Philos. Mag.}\ }\textbf {\bibinfo {volume} {21}},\ \bibinfo
  {pages} {1049} (\bibinfo {year} {1970})}\BibitemShut {NoStop}%
\bibitem [{\citenamefont {Serra}\ \emph {et~al.}(1991)\citenamefont {Serra},
  \citenamefont {Pond},\ and\ \citenamefont {Bacon}}]{Serra1991}%
  \BibitemOpen
  \bibfield  {author} {\bibinfo {author} {\bibfnamefont {A.}~\bibnamefont
  {Serra}}, \bibinfo {author} {\bibfnamefont {R.~C.}\ \bibnamefont {Pond}}, \
  and\ \bibinfo {author} {\bibfnamefont {D.~J.}\ \bibnamefont {Bacon}},\ }\href
  {\doibase 10.1016/0956-7151(91)90232-P} {\bibfield  {journal} {\bibinfo
  {journal} {Acta Metall. Mater.}\ }\textbf {\bibinfo {volume} {39}},\ \bibinfo
  {pages} {1469} (\bibinfo {year} {1991})}\BibitemShut {NoStop}%
\bibitem [{Note1()}]{Note1}%
  \BibitemOpen
  \bibinfo {note} {See Supplemental Material at
  http://link.aps.org/supplemental/10.1103/PhysRevLett.112.075504 for movies
  showing the whole process for both pyramidal and basal slips.}\BibitemShut
  {Stop}%
\bibitem [{\citenamefont {Clouet}\ \emph {et~al.}(2009)\citenamefont {Clouet},
  \citenamefont {Ventelon},\ and\ \citenamefont {Willaime}}]{Clouet2009b}%
  \BibitemOpen
  \bibfield  {author} {\bibinfo {author} {\bibfnamefont {E.}~\bibnamefont
  {Clouet}}, \bibinfo {author} {\bibfnamefont {L.}~\bibnamefont {Ventelon}}, \
  and\ \bibinfo {author} {\bibfnamefont {F.}~\bibnamefont {Willaime}},\ }\href
  {\doibase 10.1103/PhysRevLett.102.055502} {\bibfield  {journal} {\bibinfo
  {journal} {Phys. Rev. Lett.}\ }\textbf {\bibinfo {volume} {102}},\ \bibinfo
  {pages} {055502} (\bibinfo {year} {2009})}\BibitemShut {NoStop}%
\end{thebibliography}%

\end{document}